\documentclass[11pt]{article}
\usepackage{amsmath,amsfonts, amssymb, braket, bbold}
\usepackage{tikz}
\usetikzlibrary{trees,er,snakes,shapes,mindmap}
\usepackage{titlesec}
\titleformat{\section}
 {\normalfont\fontfamily{put}\fontsize{12pt}{16pt}\bfseries\color{black}}
{\thesection}{1em}{}
\def \beq  {\begin{equation}}
\def \eeq  {\end{equation}}
\def \beqar {\begin{eqnarray}}
\def \eeqar {\end{eqnarray}}
\allowdisplaybreaks
\def\sqr#1#2{{\vcenter{\vbox{\hrule height.#2pt
\hbox{\vrule width.#2pt height#1pt \kern#1pt
\vrule width.#2pt}\hrule height.#2pt}}}}

\def\vf {{\varphi}}

\def\Tr {{\rm Tr}}

\def\del {\partial}

\def\bz {{\bar{z}}}

\def\C {{\cal C}}
\def\D {{\cal D}}

\def\H {{\cal H}}

\def\M{{\cal M}}

\def\vf {{\varphi}}

\def\half{\textstyle{1\over 2}}

\mathchardef\mhyphen="2D
\begin{document}
\fontfamily{put}\fontsize{11pt}{16pt}\selectfont
\def \CMP {{Commun. Math. Phys.}}
\def \PRL {{Phys. Rev. Lett.}}
\def \PL {{Phys. Lett.}}
\def \NPBProc {{Nucl. Phys. B (Proc. Suppl.)}}
\def \NP {{Nucl. Phys.}}
\def \RMP {{Rev. Mod. Phys.}}
\def \JGP {{J. Geom. Phys.}}
\def \CQG {{Class. Quant. Grav.}}
\def \MPL {{Mod. Phys. Lett.}}
\def \IJMP {{ Int. J. Mod. Phys.}}
\def \JHEP {{JHEP}}
\def \PR {{Phys. Rev.}}
\def \JMP {{J. Math. Phys.}}
\def \GRG{{Gen. Rel. Grav.}}
\begin{titlepage}
\null\vspace{-62pt} \pagestyle{empty}
\begin{center}
\vspace{1.3truein} {\Large\bfseries
Landau-Hall States and Berezin-Toeplitz Quantization of}\\
\vskip .15in
{\Large\bfseries Matrix Algebras}\\
\vskip .5in
{\Large\bfseries ~}\\
{\large\sc V.P. Nair}\\
\vskip .2in
{\itshape Physics Department,
City College of the CUNY\\
New York, NY 10031}\\
 \vskip .1in
\begin{tabular}{r l}
E-mail:&\!\!\!{\fontfamily{cmtt}\fontsize{11pt}{15pt}\selectfont vpnair@ccny.cuny.edu}\\
\end{tabular}
\vskip 1.5in
\centerline{\large\bf Abstract}
\end{center}
We argue that the Landau-Hall states provide a suitable framework for 
formulating the Berezin-Toeplitz quantization of classical functions on a K\"ahler 
phase space. We derive the star-products for such functions in this framework
and generalize the Berezin-Toeplitz quantization to matrix-valued classical
functions. We also comment on how this is related to different calculations 
of the effective action for Hall systems.

\end{titlepage}
\pagestyle{plain} \setcounter{page}{2}
\section{Introduction}
In constructing a physical theory, as the basic postulate we must start with the full quantum theory, obtaining the classical theory as a suitable approximation for certain regimes of parameters.
But the {\it a priori} deduction of the quantum version of the theory from experimental data is quite difficult, mainly because we obtain data using classical apparatus and
hence our intuition is largely based on classical physics. So we have the process
of quantization whereby classical observables, i.e., functions on a phase space, are mapped 
to self-adjoint operators on a Hilbert space. There are many quantization approaches developed to deal with the inherent ambiguities associated with this mapping due to operator ordering issues, self-adjointness problems, etc. These include various correspondence principles, geometric
quantization, deformation quantization, etc. The Berezin-Toeplitz quantization
\cite{ber}-\cite{schlich} is a procedure for mapping a classical function $A(z, \bz )$ to an operator by defining the matrix elements 
$A_{ij}$ of the corresponding operator ${\hat A}$ as
\beq
A_{ij} = \int_{\M} d V~ \Psi^*_i \, A(z, \bz ) \, \Psi_j
\label{intr1}
\eeq
Here $\M$ is a complex K\"ahler manifold with complex coordinates $z^\alpha, \bz^\alpha$.
$\Psi_i$ are a complete set of coherent state wave functions on $\M$ satisfying a holomorphicity
condition, and $d V$ is the volume element for $\M$. The function
$A(z, \bz )$ is known as the contravariant symbol for
the operator ${\hat A}$. As an example, if we consider a K\"ahler manifold of the coset type
$\M = G/H$ for a compact Lie group $G$, with $H$ being a suitable subgroup,
the coherent states are of the form
\beq
\Psi_k = \sqrt{{\rm dim}J} \, \bra{J, k} g \ket{J, w}
\label{intr2}
\eeq
where $\bra{J, k} g \ket{J, l}$ denotes the $(k,l)$-matrix element of the group
element $g$ in a representation denoted as $J$. The state $\ket{J, w}$ is to be chosen
as a state (or set of states) carrying a specific representation of $H$.

Equation (\ref{intr1}) describes the transition from a classical function to an operator.
The converse question is to obtain a classical function given an operator. This is done by the
covariant symbol, which can be defined
as
\beq
(A) = \C \sum_{k,l} \Psi_k \, A_{kl} \Psi^*_l
\label{intr3}
\eeq
where $\C$ is a factor depending on normalizations. For our example of
$\M= G/H$, $\C = (1/ {\rm dim}J)$.

The contravariant and covariant symbols are not exact inverses in the sense that
if we start from $A(z, \bz)$, construct $A_{kl}$ using (\ref{intr1}) and then
use (\ref{intr2}), the $(A)$ so obtained is not $A(z, \bz)$. 
The exact inverse process would be to identify a function $A(z, \bz)$ such that
(\ref{intr1}) holds where we are given $A_{kl}$ as the input information.
The answer to this is the diagonal coherent state representation \cite{sudar}.

There are a couple of questions which arise
naturally given this layout of BT quantization.
The first is: How do we define a star product which realizes the operator algebra at the
level of the contravariant symbols?
Explicit formulae for the star product have been obtained before (see
\cite{schlich} for a review), but 
we will argue that an easier approach involves considering the coherent states
as corresponding to the lowest Landau levels of a quantum Hall problem
\cite{KN-1}-\cite{KN-rev}.
This embedding of the problem in the larger framework
transforms it to a field theory problem and gives a simple way to write the
star product\footnote{The Landau-Hall framework is close to the formulation of fuzzy spaces.
For a discussion of BT quantization as applied to matrix models relevant for
M-theory, see \cite{ishiki}.}.

The second question we might ask is about the BT quantization
of matrix-valued classical functions. There are situations where such functions,
with a noncommutative matrix algebra, can arise already at the classical level.
(Defining a classical field theory with nonabelian symmetries
on a noncommutative space would be one example.)
Framing this question as a Landau problem, we can give a definition and construct
the corresponding star product.  The latter reduces to the matrix algebra
at the lowest order, as expected. 

There is another somewhat nuanced issue on the physics side of things
which is clarified by this work.
For the quantum Hall states one can define an effective action in terms of 
the external gauge fields which is obtained by integrating out the fermion fields
\cite{QHE}. This is standard procedure
in the field theory and involves virtual transitions between the lowest Landau level
and the higher levels \cite{RS}. On the other hand, one can just consider the subspace of
states in the lowest Landau level
and calculate an effective action \cite{KN-1}. This would involve the use of
covariant symbols for operators. The embedding of BT quantization in the framework
of the Landau problem shows that the first procedure is identical to the use of the
contravariant symbol, clarifying the relation between these two approaches.

What is outlined in the previous three paragraphs summarize the key results of this paper.
As for the rest of this paper, in section 2, we consider the case of $S^2$, 
construct the star products and show consistency with the expected asymptotic 
behavior. In section 3, we do the analysis for matrix-valued functions.
In the discussion, we make more specific comments on the relevance to the
calculation of the effective action for the Hall problem.

\section{Landau problem on $S^2$}

We consider the two-sphere $S^2$ as a complex manifold ${\mathbb{CP}}^1$. It can also be considered as $SU(2)/U(1)$. This tells us that the Riemann curvature tensor of $S^2$ takes values in
the Lie algebra of $U(1)$ and that it is constant in a suitable choice of frames.
One can then consider an additional background $U(1)$ field which is proportional to the curvature and hence is constant on $S^2$.
Such a field would be like the magnetic field of a magnetic monopole sitting at the center if we consider the $S^2$ as embedded in ${\mathbb R}^3$ in the usual way.
The Landau problem refers to the dynamics of a charged particle in such a background field
\cite{{haldane}, {KN-1}}.
It is described by the Hamiltonian
\beq
H = - {\D^2 \over 2 m }
\label{BT1}
\eeq
where $\D_i$ is the covariant derivative on the sphere
in the background of the constant magnetic field, $\D^2$ being the covariant Laplacian.
($m$ is the mass of the particle.)
This can be phrased in terms of the generators of the group $SU(2)$. Translation operators
on the sphere in complex coordinates correspond to the group generators
$R_\pm$, which obey the $SU(2)$ Lie algebra relations
\beq
[R_+, R_- ] = 2 \, R_3, \hskip .3in
[R_3, R_\pm ] = \pm \, R_\pm
\label{BT2}
\eeq
where the third generator $R_3$ corresponds to the $U(1)$ generator. The Hamiltonian takes the form
\beq
H = {R_+ R_- + R_- R_+ \over 4 m r^2} 
\label{BT3}
\eeq
the identification being $\D_\pm = i R_\pm/r $. Here $r$ is a scale factor corresponding to the radius of the sphere.
Since $[D_+, D_- ] =  2 B$, where $B$ is the magnetic field,
we see, by comparing this to the commutator for $R_\pm$, that
we need $R_3 = - n/2$ on the states of interest, where $n = 2 B r^2$.
The fact that $n$ must be an integer to obtain unitary representations of $SU(2)$ is just the
standard Dirac quantization condition.

The eigenfunctions of the Hamiltonian are then easy to construct.
Let $g$ denote an element of the group $SU(2)$ in the fundamental representation,
as a $2\times 2$  matrix. Explicitly, on a coordinate patch, we can parametrize
$g$ as
\beq
g = {1\over \sqrt{1+ \bz z}} \left( \begin{matrix} \bz & 1\\
-1&z\\ \end{matrix} \right)\, \left( \begin{matrix} e^{i\vf/2} &0\\ 0& e^{-i \vf/2} \\ \end{matrix}
\right)
\label{BT3a}
\eeq
corresponding to complex coordinates $z, \bz$ for one coordinate patch on $S^2$, they are the coordinates defined by a stereographic projection of $S^2$;
$\vf$ is
the angular parameter for which $R_3$ is the translation operator.

We can define the left and right action  of the group
generators on $g$ by
\beq
L_a \, g = t_a\, g, \hskip .3in
R_a \, g = g\, t_a, \hskip .2in a= 1, 2, 3,
\label{BT4}
\eeq
where $t_a = {\half} \sigma_a$, $\sigma_a$ being the Pauli matrices.
The eigenfunctions of the Hamiltonian can be obtained in terms of the representative
of $g$ in an arbitrary representation where the eigenvalue for the action of $R_3$ 
is fixed to be $- {n \over 2}$. Explicitly, they
are given by
\beqar
\Psi^{(q)}_{k} &=& \sqrt{ n + 2 q+ 1}\,D^{(q)}_{k, -{n \over 2}}\nonumber\\
D^{(q)}_{k, -{n \over 2}} &=& \bra{{\textstyle{n\over 2}}+ q, k} \, {\hat g} \ket{{\textstyle{n\over 2}} +q, -{\textstyle{n\over 2}}}
\label{BT5}
\eeqar
in terms of the standard notation for $SU(2)$ eigenstates as $\ket{j, m}$. Here
$j = {n \over 2} + q$;
$q$ is a positive semi-definite integer taking values $0, 1, 2, $ etc., corresponding to the various energy levels.
These wave functions are $L^2$-normalized with the standard Haar measure on the group,
\beq
\int d\mu \, \Psi^{(q)*}_{k} \Psi^{(q)}_{l} = \delta_{kl}
\label{BT5a}
\eeq
where $d\mu $ is the volume element for the group.
For these states, it is also easy to see that 
\beq
R_+ R_- \, \Psi^{(q)}_k = (R^2 - R_3^2 + R_3 ) \, \Psi^{(q)}_k = ( q n + q (q+1)) \, \Psi^{(q)}_k
\label{BT6}
\eeq
Notice that the group generators $L_a$ for the left action
defined by (\ref{BT4}) commute with the Hamiltonian and so
the states in (\ref{BT5}) are degenerate for all values of $k$ for a given 
$j = {n \over 2} + q$. The degeneracy is therefore
$ 2j +1 = n + 2 q + 1$, for a given $q$.
We may also note that $\Psi^{(q)}_k$ are not, properly speaking, functions on
$S^2$, since they have a nontrivial transformation under
$R_3$. They are sections of a $U(1)$-bundle over $S^2$.

It is useful to look at the lowest states in some detail.
These correspond to $q = 0$, so $- {n \over 2}$ is the lowest possible
value for $m$, in the $\ket{j,m}$ notation, since $j = {n \over 2}$.
We thus have
\beq
R_- \, \Psi^{(0)}_k  = 0
\label{BT7}
\eeq
This is a holomorphicity condition on the lowest set of eigenfunctions, corresponding to
$q = 0$.
We may regard $\Psi^{(0)}_k$ as wave functions corresponding to
coherent states for the two-sphere. They can be
obtained by straightforward geometric quantization of the canonical structure
\cite{geom}
\beq
\Omega = n\, \omega = {n \over 2}\, \left[ -i \Tr (\sigma_3 g^{-1} d g \, g^{-1} dg )
\right]
\label{BT8}
\eeq
Here $\omega$ is the K\"ahler two-form on $S^2= SU(2)/U(1)$.
Thus $\Psi^{(0)}_k$ correspond to sections of the $n$-th power of the canonical line bundle
of $SU(2)/U(1)$.
These lowest levels could be obtained, via geometric quantization of
(\ref{BT8}), without the need for the full set of eigenfunctions
for the Hamiltonian (\ref{BT3}). 
They form an orthonormal basis for a Hilbert space $\H_0$ of dimension $n+1$.

The Hilbert space $\H_0$ is the space of interest for us. Given a function $A (z, \bz )$
on the classical phase space $S^2$, we can define an
operator ${\hat A}$ acting on $\H_0$ by its matrix elements as
\beq
A_{kl} \equiv \bra{k} {\hat A} \ket{l} =
\int d\mu \, \Psi^{(0)*}_k \, A(z, \bz) \Psi^{(0)}_l
\label{BT10}
\eeq
This correspondence of assigning an operator to the function $A(z, \bz)$
is the Berezin-Toeplitz (BT) quantization of $A(z, \bz)$. If ${\hat A}$ is taken as
the given quantity, the function $A(z, \bz )$ which leads to it
via (\ref{BT10}) is referred to as the contravariant symbol
for ${\hat A}$.
Starting from the operator, one can also define a covariant symbol which is a function
on the phase space $S^2$, given by
\beq
(A) = D^{(0)}_{k, -{n\over 2}} \, A_{kl} \, D^{(0)*}_{l, -{n \over 2}}
\label{BT11}
\eeq
Notice that the  group elements are used in this definition, not the correctly normalized
wave functions.

The contravariant and covariant symbols are converses of each other albeit in a
qualified sense: 
In the first case we start with the function $A(z,\bz)$ on $S^2$ and define the associated
operator via (\ref{BT10}). In the second case, we start with the matrix elements of the
operator and define a function, namely $(A)$, on $S^2$.
If we start from $A(z,\bz)$ and define the operator and take its covariant symbol, in general,
we do not get back the function $A(z, \bz)$. The map $A(z, \bz ) \rightarrow (A)$ is
the Berezin transform \cite{schlich}.
We will see shortly that as $n \rightarrow \infty$, $(A) $ becomes the same as
$A(z, \bz)$.

The construction of star products for the covariant symbol is fairly simple \cite{{KN-1},{KN-rev}}.
The symbol for the product of two operators, say $A$ and $B$, is, by definition,
\beq
(AB) = D^{(0)}_{k, -{n\over 2}} \, A_{kl} B_{lp} D^{(0)*}_{p, -{n\over 2}}
\label{BT12}
\eeq
We can insert $\delta_{ml} = D^{(0)*}_{m, r} \,D^{(0)}_{l, r}$
into this expression and simplify it as follows.
\beqar
(AB) &=& D^{(0)}_{k, -{n\over 2}} \, A_{km} \, D^{(0)*}_{m, r} \,D^{(0)}_{l, r}\,
B_{lp} D^{(0)*}_{p, -{n\over 2}}\nonumber\\
&=& D^{(0)}_{k, -{n\over 2}} \, A_{km} \, D^{(0)*}_{m, -{n\over 2} }\,D^{(0)}_{l, -{n \over 2}}\,
B_{lp} D^{(0)*}_{p, -{n\over 2}}\nonumber\\
&&\hskip .2in
+ D^{(0)}_{k, -{n\over 2}} \, A_{km} \, D^{(0)*}_{m, - {n\over 2}+1} \,D^{(0)}_{l, -{n \over 2}+1}\,
B_{lp} D^{(0)*}_{p, -{n\over 2}} + \cdots\nonumber\\
&=& (A) (B) - {1\over n} D^{(0)}_{k, -{n\over 2}} \, A_{km} \, R_- D^{(0)*}_{m, - {n\over 2}} \,R_+ D^{(0)}_{l, -{n \over 2}}\,B_{lp} D^{(0)*}_{p, -{n\over 2}} + \cdots\nonumber\\
&=& (A) (B) - {1\over n} R_- (A) \, R_+(B) + \cdots\nonumber\\
&=& (A)*(B)
\label{BT13}
\eeqar
where we have used
\beq
D^{(0)}_{l, -{n \over 2}+1} = {1\over \sqrt{n}} R_+ D^{(0)}_{l, -{n \over 2}},
\hskip .2in
D^{(0)*}_{m, -{n \over 2}+1} = - {1\over \sqrt{n}} R_- D^{(0)*}_{m, -{n \over 2}}
\label{BT14}
\eeq
We have also used the fact that, since $R_-  D^{(0)}_{k, -{n\over 2}} = 0$, we can write
\beq
D^{(0)}_{k, -{n\over 2}} A_{km} R_- D^{(0)*}_{m, - {n\over 2}}
= R_- \left( D^{(0)}_{k, -{n\over 2}} A_{km} D^{(0)*}_{m, - {n\over 2}}\right)
= R_- (A)
\label{BT16}
\eeq
with a similar simplification for the symbol for $B$. It is clear that the higher terms
in (\ref{BT13}) can be simplified in a similar way and written in terms
of $R_-^s(A) \, R_+^s(B)$ for $s> 1$. For the covariant symbol as we have defined it,
the series terminates, for 
finite $n$.

The construction of a star product for the contravariant symbol is more involved.
Here the question is to find a product of the classical functions $A(z, \bz)$ and 
$B(z, \bz )$ such that the BT quantization of the product gives the product of
the operators.
The product of the BT quantized operators is given by
\beqar
A_{kl} B_{lm} &=& \int d\mu \, \Psi^{(0)*}_k \, A(z, \bz) \Psi^{(0)}_l\,
\int d\mu' \, \Psi^{(0)*}_l \, B(z', \bz') \Psi^{(0)}_m\nonumber\\
&=& \int d\mu d\mu'\, \, \Psi^{(0)*}_k \, A(z, \bz) \, P^{(0)}(g,g') \, B(z', \bz') \Psi^{(0)}_m
\label{BT17}
\eeqar
where $P^{(0)}(g,g')$ is the projection operator for the lowest Landau level,
\beq
P^{(0)}(g,g') = \sum_l \Psi^{(0)}_l (g)\,  \Psi^{(0)*}_l (g')
\label{BT18}
\eeq
We denote the arguments as $g$ and $g'$ for brevity,
although $P^{(0)}$ is defined
on the coset $SU(2)/U(1)$, i.e., independent of the $U(1)$ angle $\vf$.
If we consider similar projection operators to the higher levels, we have, by completeness
of all the eigenstates of the Hamiltonian (\ref{BT1}),
\beq
\delta(g,g') = \sum \Psi^{(q)}_k (g)\,  \Psi^{(q)*}_k (g')
= \sum_q P^{(q)}(g,g')  = P^{(0)} (g,g') + P^{(1)}(g, g') + \cdots
\label{BT19}
\eeq
Our strategy will be to write $P^{(0)} (g,g')$ in terms of
$\delta(g,g')$ and derivatives acting on $\delta(g,g')$.
Notice that the action of $R_+ R_- $ on $P^{(0)}(g,g') $ is zero,
due to (\ref{BT7}).
We also have the result
\beq
R_+ R_- \Psi^{(q)}_k (g) = \bigl(q n + q (q+1)\bigr) \, \Psi^{(q)}_k (g)
\label{BT20}
\eeq
Using this result, we can eliminate $P^{(1)}(g, g')$ from 
(\ref{BT19}) and write
\beq
\delta (g, g') - {1\over (n+2)} R_+ R_- \delta (g, g') 
= P^{(0)} (g, g') + \sum_{q=2}^\infty  \left[ 1 - {q n + q (q+1) \over (n+2)}\right]
P^{(q)}(g, g') 
\label{BT21}
\eeq
If we use this result
for $P^{(0)}(g, g')$ in (\ref{BT17}) we can
write $ A(z, \bz) \, P^{(0)}(g,g') \, B(z', \bz')$ in terms of
$ A(z, \bz) \,  B(z', \bz')$ and products of derivatives of these functions, as we would expect for
a star product.
This is the basic idea.
To carry this out to higher orders, we need to eliminate
$P^{(q)}$, $q \geq 2$, at least recursively. Therefore,
more generally, we start by writing $P^{(0)}$ as
\beq
P^{(0)} (g, g' ) 
= \Bigl[ 1 + \sum_s c_s\, R^s_+ R^s_- \Bigr] \, \delta(g, g')
\label{BT22}
\eeq
for some constant coefficients $c_s$.
The key property we need  is that $R^s_- \Psi^{(q)}_k(g) = 0$ for $s > q$, so we can recursively define
the coefficients $c_s$ to eliminate the contribution of the higher levels
in $\delta (g, g') =  \sum \Psi^{(q)}_k (g)\,  \Psi^{(q)*}_k (g')$.
The conditions we need are
\beq
 \Bigl[ 1 +  \sum_{s=1}^q c_s\, R^s_+ R^s_- \Bigr] \, \Psi^{(q)}_k (g) = 0
 \label{BT23}
 \eeq
 It is easy to work out the action of $R^s_+ R^s_-$ on the wave functions,
 \beqar
 R^s_+ R^s_- \, \Psi^{(q)}_k (g) &=& \left[\prod_1^s
f(q,s)\right] \, \Psi^{(q)}_k (g)\nonumber\\
f(q,s) &=&  \left[ n (q-s+1) + q (q+1) - s (s-1) \right]
\label{BT24}
\eeqar
One can recursively calculate $c_s$. The lowest two coefficients are
\beq
c_1 = - {1\over f(1,1)} = - {1\over (n+2)},
\hskip .2in
c_2 = {1\over f(2,2)} \left( {1\over f(1,1)} - {1\over f(2,1)} \right) = 
{1\over 2 (n+2) (n+3)}
\label{BT25}
\eeq
Using (\ref{BT22}) in (\ref{BT17}), we get
\beqar
A_{kl} B_{lm}
&=& \int d\mu d\mu'\, \, \Psi^{(0)*}_k \, A(z, \bz) \, P^{(0)}(g,g') \, B(z', \bz') \Psi^{(0)}_m
\nonumber\\
&=& \int d\mu \, \, \Psi^{(0)*}_k \, A(z, \bz) \left[ 1
+ \sum_s c_s R^s_+ R^s_- \right] \, B(z', \bz') \Psi^{(0)}_m\nonumber\\
&=& \int d\mu \, \Psi^{(0)*}_k \, A(z, \bz)\, B(z,\bz ) \Psi^{(0)}_m\nonumber\\
&&\hskip .2in+ \int d\mu \sum_s (-1)^s c_s  \bigl(R^s_+ \Psi^{(0)*}_k A(z, \bz)\bigr)
\bigl( R^s_- \, B(z', \bz') \Psi^{(0)}_m\bigr)\nonumber\\
&=&  \int d\mu \, \Psi^{(0)*}_k \, A(z, \bz)\, B(z,\bz ) \Psi^{(0)}_m\nonumber\\
&&\hskip .2in+ \int d\mu \sum_s (-1)^s c_s  \Psi^{(0)*}_k \bigl(R^s_+ A(z, \bz)\bigr)
\bigl( R^s_- \, B(z', \bz') \bigr) \Psi^{(0)}_m\nonumber\\
&=&  \int d\mu \, \Psi^{(0)*}_k \, \bigl[ A(z, \bz)* B(z,\bz )\bigr] \Psi^{(0)}_m
\label{BT26}
\eeqar
\beq
A(z, \bz)* B(z,\bz ) =
A(z, \bz) \, B(z,\bz ) + \sum_{s=1}^\infty
 (-1)^s c_s  \bigl(R^s_+ A(z, \bz)\bigr)
\bigl( R^s_- \, B(z', \bz') \bigr)
\label{BT27}
\eeq
In (\ref{BT26}), in the second step, we did integration by parts to move
$R^s_+$ to act on $ \Psi^{(0)*}_k A(z, \bz)$ and then used the fact that
$R_+ \Psi^{(0)*}_k  = 0$. In this way we are able to isolate the product of functions
which, upon BT quantization, reproduces the operator product.
Notice that this is an infinite series even for finite $n$, unlike the star product we defined
on the covariant symbols. The first two terms of this star product can be written out, using
(\ref{BT25}), as
\beq
A * B =
A \, B  + {1\over (n+2)} (R_+ A) \, (R_- B) + {1\over 2 (n+2) (n+3)}
(R^2_+ A) \, (R^2_-B) + \cdots
\label{BT28}
\eeq
This star product is different from what we found for the covariant symbol.
This is to be expected since the covariant symbol is different from
the contravariant symbol. However, notice that, in the large $n$ limit,
\beqar
(A)*(B) - (B)*(A) &=& {1\over n} \left[ R_+ (A) \, R_- (B) - R_- (A). R_+ (B) \right] + {\cal O}(1/n^2)\nonumber\\
&=& A*B - B*A + {\cal O} (1/n^2)
\label{BT29}
\eeqar
The Poisson bracket is recovered for the star commutator for
both cases in the large $n$ limit.

We now return to the relation between the contravariant and covariant
symbols. The Berezin transform for $A$ is given by
\beqar
(A) &=& D^{(0)}_{k, -{n\over 2}} \, \Bigl[\int d\mu' \Psi^{(0)*} (g') A(z',\bz')\Psi^{(0)}_l(g')\Bigr]
 D^{(0)*}_{l, -{n\over 2}}
 \nonumber\\
 &=&\int d\mu' \, P^{(0)}(g,g') A(z', \bz') \, D^{(0)}_{k, -{n\over 2}}(g') D^{(0)*}_{k, -{n\over 2}}(g)
 \nonumber\\
 &=&A(z, \bz) + \sum_s c_s R^s_+ R^s_- \Bigl( A D^{(0)}_{k, -{n\over 2}}(g)\Bigr)
 D^{(0)*}_{k, -{n\over 2}}(g)\nonumber\\
 &=&A(z, \bz) + \sum_s c_s R^s_+  \Bigl((R^s_- A) \,D^{(0)}_{k, -{n\over 2}}(g)\Bigr)
 D^{(0)*}_{k, -{n\over 2}}(g)\nonumber\\
 &=&A(z, \bz) + \sum_s c_s (R^s_+ R^s_- A) \,D^{(0)}_{k, -{n\over 2}}(g)
 D^{(0)*}_{k, -{n\over 2}}(g)\nonumber\\
 &=&A(z, \bz) + \sum_s c_s (R^s_+ R^s_- A)
 \label{BT30}
 \eeqar
 We have used (\ref{BT22}) to obtain the third line
 of this equation, and $R_- D^{(0)}_{k, -{n\over 2}}(g) = 0$ to move to the fourth line.
 Further, we have
 $D^{(0)}_{k, -{n\over 2}}(g)
 D^{(0)*}_{k, -{n\over 2}}(g) = 1$ by the group property and 
 \beq
\sum_k \left[ R^s_+ D^{(0)}_{k, -{n\over 2}}(g) \right] D^{(0)*}_{k, -{n\over 2}}(g)
\sim D^{(0)}_{k, -{n\over 2}+ s}(g) D^{(0)*}_{k, -{n\over 2}}(g)
\sim \braket{-\textstyle{{n\over 2}}|- \textstyle{{n\over 2}}+s}
 = 0
 \label{BT31}
 \eeq
 This was used in the last two steps. 
 We see that, in the large $n$ limit, the two symbols will coincide. Further,
 \beqar
 (A) &=& A - {1\over (n+2)} R_+ R_- A + \cdots\nonumber\\
&=& A + {1\over (n+2)} \Delta A + \cdots
 \label{BT32}
 \eeqar
 where $\Delta$ is the Laplace operator on $S^2$.
 This result is consistent with the theorem of Karabegov and Schlichenmaier,
 quoted as theorem 7.2 in \cite{schlich}, although our derivation is very different.

\section{BT quantization for matrix-valued functions}

In the last section, the wave functions we used were those for the Landau problem
on $S^2$ with an Abelian $U(1)$ background field.
In other words, they were sections of an appropriate line bundle.
The lowest set of such wave functions, which were sections of a holomorphic
line bundle, were then used to define the BT quantization of a function
on $S^2$.
Generalizing, we can use the sections of a suitably chosen vector bundle to define the BT quantization
of a matrix-valued function. Naturally, this will mean using
a background field corresponding to a nonabelian group
${H}$, the wave functions being those of a particle transforming nontrivially according to
some representation of ${H}$ \cite{KN-1}. 
We will now work out an example of how this can be done using
${\mathbb{CP}}^2 = SU(3)/U(2)$ as the manifold of interest.
The group translation operators can be separated into
$R_a$, $a = 1,2,3$, which are generators of $SU(2) \in U(2) \in SU(3)$,
$R_8$ which is the generator for $U(1) \in U(2)$ and
the remaining coset generators $R_{+i}, \, R_{-i}$, $i = 1, 2$.
The last set $R_{\pm i}$ are the translation generators for
${\mathbb{CP}}^2$, with the Hamiltonian taken as
\beq
H = {R_{+i} R_{-i} \over 2 m r^2}
\label{BT33}
\eeq
The eigenfunctions are given by
\beq
\Psi_{A; a} = C\, \bra{A} {\hat g} \ket{\alpha}
\label{BT34}
\eeq
$C$ is a normalization constant, we will discuss this below.
The index $A$ labels states in the $SU(3)$ representation to be specified
by the choice of $\ket{\alpha}$.
We will consider a combination of $U(1)$ and $SU(2)$ background fields.
Thus the state $\ket{\alpha}$ must be chosen so that it transforms according
to the required representation of $SU(2)$
(equivalent to specifying the $SU(2)$ charges) and also
carries the required $U(1)$ charge. This means
\beqar
R_a \ket{\alpha} &=& (T_a)_{\alpha \beta} \, \ket{\beta}\nonumber\\
R_8 \ket{\alpha}&=& - {n \over \sqrt{3}} \ket{\alpha}
\label{BT35}
\eeqar
$(T_a)_{\alpha \beta}$ are the charge matrices for the coupling of the
particle to the constant $SU(2)$ background field.
In the tensor notation $T^P_Q$ for $SU(3)$ representations with
$P$ up-indices and $Q$ down-indices (where each index can take values $1, 2, 3$),
the possible choices for the states $\ket{\alpha}$ are then of the form
\beq
\ket{\alpha} = \ket{P; Q} = \ket{33\cdots 3 ;
33\cdots 3 ; i_1 \cdots i_l}
\label{BT36}
\eeq
This corresponds to 
$q$ down-indices all set to $3$, with an additional $l$ indices, each of which can take
values $1, 2$ corresponding to the $SU(2)$ spinors.
There are also $n - (l/2) + q$ up-indices all set to $3$.
Thus
this state transforms as the spin-$l/2$ representation of $SU(2)$, i.e.,
$(T_a)_{\alpha \beta}$ are $(l+1)\times (l+1) $ matrices, and has 
the $R_8$ value
\beq
R_8  \ket{33\cdots 3 ;
33\cdots 3 ; i_1 \cdots i_l} = - {1\over \sqrt{3}}  (n - {\half} l + q)  +{1\over \sqrt{3}} q 
- {l \over 2 \sqrt{3}} = - {n \over \sqrt{3}}
\label{BT37}
\eeq
as required in (\ref{BT35}).  Notice that $l$ should be an even integer, to get an integer number of
up-indices, so that we must have integer spin representations for $SU(2)$.
(Ultimately, this is related to the fact that  ${\mathbb{CP}}^2$ does not admit a spin structure.)
The eigenvalues of $R_{+i} R_{-i}$ are given by
\beqar
R_{+i} R_{-i} \Psi^{(q)}_{A; \alpha} &=&
\left[ C_2(SU(3)) - C_2(SU(2) - R_8^2 + \sqrt{3} R_8 \right]\, \Psi^{(q)}_{A; \alpha}\nonumber\\
&=&\left[ q n + q (q +2 + {\half}l )\right]\, \Psi^{(q)}_{A; \alpha}
\label{BT38}
\eeqar
The lowest state corresponding to $ q= 0$ obeys the
holomorphicity condition
$R_{-i} \Psi^{(0)}_{A; \alpha} = 0$.
The degeneracy of the states as given by the dimension of
the representation is
\beq
N = {1\over 2} \left(n +q + 1 - {l \over 2} \right)\, (q + l +1) \,\left( n + 2 q + {l \over 2} + 2 \right)
\label{BT39}
\eeq

We now consider the normalization of these states. Wave functions such as 
(\ref{BT34}) arise for fields which carry the spin-${l \over 2}$ representation of
$SU(2)$, so that it can be coupled to the appropriate $SU(2)$ background.
Denoting such a field by $\phi_\alpha$, the action for the field may be taken as
\beq
S = \int dt d\mu\, \left[ i \phi_\alpha^* {\del \phi_\alpha \over \del t}
- \phi_{\alpha} H_{\alpha \beta} \phi_\beta \right]
\label{BT40}
\eeq
The appropriate normalization for the one-particle wave functions should thus be
\beq
\int d\mu \, \Psi^{(q)*}_{A; \alpha} \Psi^{(q')}_{B; \alpha} 
= \delta_{AB} \delta^{qq'}
\label{BT41}
\eeq
(There is summation over $\alpha$ in this equation.) With the standard orthogonality relation
for group elements in arbitrary representations,
the normalized wave functions are
\beq
\Psi^{(q)}_{A; \alpha}  = \sqrt{N\over l+1} \, \bra{A} {\hat g} \ket{\alpha}
\label{BT42}
\eeq
The completeness relation for these states is
\beq
\sum_{q, A} \Psi^{(q)}_{A; \alpha} (g) \, \Psi^{(q)*}_{A; \beta}(g')  =
\delta (g, g')\, \delta_{\alpha\beta}
\label{BT43}
\eeq

We are now in a position to write down the BT quantization of a matrix-valued function
$A_{\alpha \beta}$ as
\beq
A_{CD} = \int d\mu \, \Psi^{(0)*}_{C; \alpha}(g) \, A_{\alpha \beta}\,
\Psi^{(0)}_{D; \beta}(g) 
\label{BT44}
\eeq
The integrand should be a function defined on ${\mathbb{CP}}^2$
and hence invariant under the $U(2)$ subgroup for the integral to 
be nonzero. This means that the nontrivial transformation
of $\Psi^{(0)*}_{C; \alpha}(g)$ and $\Psi^{(0)}_{D; \beta}(g) $ should be 
compensated by a suitable transformation of $A_{\alpha\beta}$.
Thus strictly speaking, already at the classical level,
 we are not considering matrix-valued functions, but sections of 
a suitable vector bundle.

The product of two operators defined as in (\ref{BT44})  is given by
\beqar
(AB)_{CF} &=& A_{CD} B_{DF} =
\int d\mu d\mu' \, \Psi^{(0)*}_{C; \alpha}(g) \, A_{\alpha \beta}(g)\,
\Psi^{(0)}_{D; \beta}(g) \Psi^{(0)*}_{D; \gamma}(g') \, B_{\gamma \delta}(g')\,
\Psi^{(0)}_{F; \delta}(g')\nonumber\\
&=&\int d\mu d\mu' \, \Psi^{(0)*}_{C; \alpha}(g) \, A_{\alpha \beta}(g)\,
P^{(0)}_{\beta; \gamma}(g,g') \, B_{\gamma \delta}(g')\,
\Psi^{(0)}_{F; \delta}(g')
\label{BT45}\\
P^{(0)}_{\beta; \gamma}(g,g') &=& \sum_D\Psi^{(0)}_{D; \beta}(g) \Psi^{(0)*}_{D; \gamma}(g')
\label{BT46}
\eeqar
As in the case of $S^2$, we will now write this projection operator in terms of the
Dirac $\delta$-function as
\beq
P^{(0)}_{\beta; \gamma}(g,g') 
= \left[ \delta_{\beta\gamma}  + \sum_s c_s (R_{+ i_1}R_{+ i_2} \cdots  R_{+i_s}\,R_{- i_s}
R_{- i_{s-1}} \cdots R_{-i_1})_{\beta\gamma}
\right] \delta(g, g')
\label{BT47}
\eeq
Acting on $\Psi^{(q)}_{B; \gamma}(g')$ and integrating over
$g'$, we see that $c_s$ should obey the conditions
\beq
\Psi^{(q)}_{B; \beta} (g) + \sum_{s=1}^q c_s (R_{+ i_1}R_{+ i_2} \cdots  R_{+i_s}\,R_{- i_s}
R_{- i_{s-1}} \cdots R_{-i_1})_{\beta\gamma}
\Psi^{(q)}_{B;\gamma} (g) = 0
\label{BT48}
\eeq
To make this more concrete, we need to evaluate the 
coefficient of $c_s$. It can be done recursively, with the result
\beq
\sum_s c_s (R_{+ i_1}R_{+ i_2} \cdots  R_{+i_s}\,R_{- i_s}
R_{- i_{s-1}} \cdots R_{-i_1})_{\beta\gamma}
\Psi^{(q)}_{B;\gamma} (g) = \prod_{k=1}^s h(q, k) \, \Psi^{(q)}_{B;\beta} (g)
\label{BT49a}
\eeq
where
\beq
h(q,k) = (q-k +1) (n + q +k + 1 + (l/2)) 
\label{BT49b}
\eeq
Thus the coefficients $c_s$ are determined by
\beq
1 + \sum_{s=1}^q c_s \prod_{k=1}^s h(q,k) = 0
\label{BT50}
\eeq
With the expansion (\ref{BT47}) for the projection operator, we
find
\beqar
(AB)_{CF}
&=&\int d\mu d\mu' \, \Psi^{(0)*}_{C; \alpha}(g) \, A_{\alpha \beta}(g)\,\Bigl[
\delta_{\beta\gamma} \delta(g,g')\nonumber\\
&&+ \sum_s c_s (R_{+ i_1}R_{+ i_2} \cdots  R_{+i_s}\,R_{- i_s}
R_{- i_{s-1}} \cdots R_{-i_1})_{\beta\gamma} \delta(g, g') \Bigr] \, B_{\gamma \delta}(g')\,
\Psi^{(0)}_{F; \delta}(g')\nonumber\\
&=&\int d\mu \, \Psi^{(0)*}_{C; \alpha}(g) \left[ A*B \right]_{\alpha\beta}
\Psi^{(0)}_{F; \beta}(g)\label{BT51}
\eeqar
We can now read off the star-product from this equation as
\beqar
 \left[ A*B \right]_{\alpha\beta}&=&
 A_{\alpha\gamma} B_{\gamma\beta}\nonumber\\
 &&+ \sum_s (-1)^s c_s\,(R_{+i_s} R_{+i_{s-1}} \cdots R_{+i_1} A )_{\alpha\gamma}\,
 (R_{-i_s} R_{-i_{s-1}} \cdots R_{-i_1} B)_{\gamma \beta}
 \label{BT52}
\eeqar
As before we have done integrations by parts to move the $R_+$s to act on $A$, and used
the fact that $R_{-i} \Psi^{(0)}_{F;\delta} = 0$,
$R_{+i} \Psi^{(0)*}_{C;\alpha} = 0$.
Since $A$ and $B$ are defined on $G = SU(3)$, i.e., they carry nonzero charges
under $U(2)$, the action of $R_{\pm i}$ should be considered 
as covariant derivatives.

It is straightforward to solve (\ref{BT50}) using (\ref{BT49b}) to write the
coefficients $c_s$ for some low values of $s$, as we did for the two-sphere.
For $q = 1, 2$, we can write out (\ref{BT50}) as
\beqar
1 + c_1 \, h(1,1) &=& 0\nonumber\\
1+ c_1 \, h(2,1) + c_2 \, h(2,1)\, h(2,2) &=& 0\label{BT53}
\eeqar
where the relevant $h(q,k)$ are given as
\beq
h(1,1) = (n+3 + {\half} l), \hskip .2in
h(2,1) = 2 (n+4 + {\half} l ), \hskip .2in
h(2,2) = (n+5 + {\half} l )
\label{BT54}
\eeq
The star-product from(\ref{BT52}) can thus be written more explicitly as
\beqar
 \left[ A*B \right]_{\alpha\beta}&=&
 A_{\alpha\gamma} B_{\gamma\beta}
+ {1\over (n+3 + {\half }l)}  \, (R_{+i} A)_{\alpha \gamma} \, (R_{-i} B)_{\gamma\beta}\nonumber\\
&& + {1 \over 2 (n + 3 +{\half} l ) (n+4 + {\half} l)}
 (R_{+i}R_{+j} A)_{\alpha \gamma} \, (R_{-j} R_{-i} B)_{\gamma\beta} + \cdots
 \label{BT55}
\eeqar

Once again, it is useful to see how this differs from what is obtained for the covariant symbol.
The latter is defined from the matrix elements of the operator as
\beq
(A)_{\alpha\beta} = \sum_{C,D} D^{(0)}_{C;\alpha}(g)
\, A_{CD} \, D^{(0)*}_{D;\alpha}(g)
\label{BT56}
\eeq
Using (\ref{BT44}), we can now write
\beqar
(A)_{\alpha \beta}&=& \sum_{C,D} D^{(0)}_{C;\alpha}(g)
\left[ \int d\mu'\, \Psi^{(0)}_{C;\gamma}(g') \,A_{\gamma \delta}(g') \, \Psi^{(0)}_{D;\delta} (g')\right] \,
D^{(0)*}_{D;\alpha}(g)\nonumber\\
&=& \int d\mu'\, P^{(0)}_{\alpha; \gamma}(g, g') \, A_{\gamma\delta}(g')
D^{(0)}_{D;\delta}(g') \,D^{(0)*}_{D;\alpha}(g)\nonumber\\
&=& A_{\alpha \beta}(g) + \sum_s c_s (R_{+ i_1}R_{+ i_2} \cdots  R_{+i_s}\,R_{- i_s}
R_{- i_{s-1}} \cdots R_{-i_1})_{\alpha\gamma}\, A_{\gamma\beta}\nonumber\\
&=& A_{\alpha \beta}(g)  - {1\over (n+3 + {\half} l)} (R_{+i} R_{-i} A )_{\alpha\beta}
+ \cdots
\label{BT57}
\eeqar
where we have used (\ref{BT47}) and the results
\beq
R_{-i} \, D_{D;\delta}^{(0)} (g) = 0, \hskip .2in
\Bigl[ R_{+ i_1}R_{+ i_2} \cdots  R_{+i_s} \, D^{(0)}_{D;\delta} (g) \Bigr]
D^{(0)*}_{D; \alpha} (g) = 0
\label{BT58}
\eeq
for reasons similar to what was stated after (\ref{BT31}).
We see that the covariant and contravariant symbols agree in the large $n$ limit.
Equation (\ref{BT57}) is our matrix generalization of the theorem 7.2 in \cite{schlich}.

The star-product for the covariant symbols was obtained in
\cite{KN-1} as
\beq
(A)_{\alpha \gamma} * (B)_{\gamma \beta} =  \biggl( (A)_{\alpha \gamma}\, (B)_{\gamma\beta} 
-{1\over n}  { R}_{-i}(A)_{\alpha\gamma}\,  {R}_{+i} (B)_{\gamma\beta}
\biggr) +{\cal O} \left( {1\over n^2}\right)
\label{BT59}
\eeq
The star-commutators for the contravariant symbols
can be obtained from (\ref{BT55}) as
\begin{align}
 \left[ A*B - B*A\right]_{\alpha\beta} =&\,
 [A\, B - B\, A]_{\alpha\beta}\nonumber\\
&+ {1\over n}  \,\Bigl[
 (R_{+i} A)_{\alpha \gamma} \, (R_{-i} B)_{\gamma\beta}
 -  (R_{+i} B)_{\alpha \gamma} \, (R_{-i} A)_{\gamma\beta}\Bigr] + \cdots \label{BT60}
 \end{align}
 The analogous commutator for the covariant symbols is given by
 (\ref{BT59}) as
\begin{align}
  \left[ (A)*(B) - (B)*(A)\right]_{\alpha\beta}=&\,
 [(A)\, (B) - (B)\, (A)]_{\alpha\beta}\nonumber\\
 &- {1\over n} \Bigl[ { R}_{-i}(A)_{\alpha\gamma}\,  {R}_{+i} (B)_{\gamma\beta}
 - { R}_{-i}(B)_{\alpha\gamma}\,  {R}_{+i} (A)_{\gamma\beta}\Bigr]  + \cdots
 \label{BT61}
\end{align}
There is a difference of the ordering of the matrix products in the second term, so
these are not identical even to first order in $1/n$.

\section{The one-particle field theory}

The BT quantization emerges in a completely natural way when we try to quantize a single-particle problem in the language of quantum field theory and then restrict to one, say the lowest, level.
Degeneracy of this level is important to obtain the matrix structure of the operator, and holomorphicity of the one-particle wave functions for this level
is important for constructing the star-products.
Thus effectively a suitable version of the Landau problem becomes the paradigm for
BT quantization. We have already mentioned the action for the
field theory for this case, namely, (\ref{BT40}).
The field operators may be expanded in terms of eigenfunctions of the
Hamiltonian as
\beqar
\phi_\alpha &=& \sum_C a^{(0)}_C \Psi^{(0)}_{C;\alpha} + \sum_{C'; q =1} a^{(q)}_{C'} \, \Psi^{(q)}_{C';\alpha}
\nonumber\\
\phi^\dagger_\alpha &=& \sum_C a^{(0)\dagger}_C \Psi^{(0)*}_{C;\alpha} + \sum_{C'; q =1} a^{(q)\dagger}_{C'} \, \Psi^{(q)*}_{C';\alpha}
\label{BT62}
\eeqar
where the second set of terms refers to the higher energy levels. $a^{(q)}_C$,
$a^{(q)\dagger}_C$ are the annihilation and creation operators for the 
particle represented by $\phi_\alpha$.

Now consider a one-particle operator ${\tilde B}$, which can be lifted to the field theory
as 
\beq
{\hat{B}} = \int dV\, \phi^\dagger \, {\tilde B} \, \phi
\label{BT63}
\eeq
${\tilde B}$ is, in general, a function of the coordinates, but it may contain derivatives
as well. Using the mode expansion for the fields, the operator ${\hat B}$ takes the
form
\beqar
{\hat B} &=& \sum_{k,l} a^{(0)\dagger}_C \, B_{CD} \, a^{(0)}_D
+ \sum_{(q, q')\neq (0,0); C, D}
\int dV\, \Psi^{(q)*}_{C} \,{\tilde B} \, \Psi^{(q')}_D~a^{(q)\dagger}_C\,a^{(q')}_D\nonumber\\
B_{CD} &=& \int dV\, \Psi^{(0)*}_C \, {\tilde B} \, \Psi^{(0)}_D
\label{BT64}
\eeqar
(Here we are using the simpler case of an Abelian background field
to illustrate the main point.)
If we consider dynamics restricted to the lowest level, i.e., higher levels with
$(q,q') \neq (0,0)$ are unoccupied, $a^{(q)}_C $ will annihilate the relevant set of
states and
${\hat B}$ effectively reduces to the first term.
Further, if the wave functions 
$\Psi^{(0)}_C$ for the lowest level define coherent states, derivatives appearing in ${\tilde B}$
can be replaced in  terms of conjugates of the holomorphic coordinates.
Thus $B_{kl}$ takes the form
\beq
B_{CD} = \int dV\, \Psi^{(0)*}_C \, B(z, \bz) \, \Psi^{(0)}_D
\label{BT65}
\eeq
for some {\it function} $B(z, \bz)$ on $\M$. 
We now see that, for one-particle states restricted to the lowest
level, the matrix elements of the field theory operator
${\hat B}$ are given by $B_{kl}$ as in (\ref{BT65}) or 
as in (\ref{intr1}).
Thus we get to the BT quantization of a function $B(z, \bz)$ on $\M$ from the field
theory defined by (\ref{BT40}).
Although we illustrated this relation with the Abelian background, it is clear that a similar
result will hold for a matrix-valued one-particle operator
${\tilde B}_{\alpha\beta}$ and a corresponding matrix-valued function
 $B_{\alpha\beta}(z, \bz)$.

The one-particle quantum mechanical calculation of the effective action
in \cite{RS} uses the
single-particle sector of the field theory we have outlined. From what is said above, 
this is equivalent to using the BT quantization for operators restricted to
the lowest Landau level, with the star-product as in
(\ref{BT28}).
The effective action calculations in
\cite{KN-1} uses the covariant symbol and the star-product as in
(\ref{BT13}).

 \bigskip

I thank Dimitra Karabali for a careful reading of the manuscript and
for useful comments.
This research was supported in part by the U.S.\ National Science
Foundation grant PHY-1820721
and by PSC-CUNY awards.
 

\end{document}